\documentclass[10pt]{article}
\usepackage{array}
\usepackage{svg}
\usepackage[style=apa, backend=biber]{biblatex}
\usepackage{booktabs}
\usepackage{listings}
\usepackage{xcolor}
\usepackage{tabularx}

\lstdefinestyle{json}{
    basicstyle=\ttfamily\small,
    keywordstyle=\color{blue},
    stringstyle=\color{red},
    numbers=left,
    numberstyle=\tiny\color{gray},
    breaklines=true,
    frame=single,
    backgroundcolor=\color{gray!5},
    showstringspaces=false
}

\makeatletter
\renewcommand\footnotesize{%
   \@setfontsize\footnotesize{8pt}{8pt}%
   \renewcommand{\baselinestretch}{1}\selectfont}
\makeatother

\usepackage[letterpaper]{geometry}
\usepackage{hicss}
\usepackage{times}
\usepackage[none]{hyphenat}
\usepackage{url}
\usepackage{latexsym}
\usepackage{minted}
\usepackage{indentfirst}
\usepackage{graphicx}
\graphicspath{{images/}}
\usepackage[
    style=apa,
  ]{biblatex}
\title{ Improving LLM Interpretability with User-Centric Chain-of-Thought Reasoning }

\author{
  Philipp Schröppel \\
  University of Ulm \\
  {\underline{philipp.schroeppel@uni.ulm.de}} 
}
\date{}

\begin{document}
\maketitle
\begin{abstract}
Advancing reasoning capabilities allow large language models (LLMs) to tackle increasingly complex problems, while reasoning traces—intermediate steps toward solutions—open up high-stakes applications by enabling human inspection of AI decision-making. However, current approaches prioritize model performance over human interpretability, limiting effective human-AI collaboration. In this study, we design and evaluate a human-centered approach that structures reasoning traces based on self-contained, verifiable steps, enabling users to independently assess and correct AI reasoning. Our approach uses XML-like tags to encode reasoning content and metadata, facilitating targeted feedback. Evaluation on mathematical reasoning tasks shows our approach maintains equivalent performance to standard Chain-of-Thought reasoning while enhancing interpretability. User studies demonstrate significant improvements in perceived usefulness and ease of use. This work advances understanding of how user-centric design of LLM outputs can better serve human collaboration needs in high-stakes AI deployments.
\end{abstract}

\subsubsection*{Keywords:}

LLM, human-AI collaboration, explainable AI, Chain-of-Thought reasoning, user-centric design.

\section{Introduction}
Large language models (LLMs) have transformed problem-solving across diverse domains, from healthcare and finance to education and e-commerce, demonstrating remarkable flexibility in performing complex tasks without task-specific training \parencite{raza2025industrial, brown_language_models_are_few_shot_learners}. Their ability to engage in sophisticated reasoning through Chain-of-Thought (CoT) approaches has made them particularly valuable for multi-step analytical tasks that require drawing inferences, manipulating knowledge representations, and justifying conclusions \parencite{wei_chain_of_thought}.
However, this widespread deployment has revealed critical limitations that constrain reliable AI adoption in real-world applications. LLMs are prone to hallucinations, producing convincing but factually incorrect information, and struggle with self-correction without external feedback \parencite{huang2024largelanguagemodelsselfcorrect, gou2024critic}. While automated verification tools exist in specialized domains like software development, most real-world applications leave humans responsible for evaluating and correcting LLM outputs. Reasoning traces—step-by-step explanations of the model's decision-making process—offer a promising approach to address this challenge, with recent work by Schelhorn et al. (2025) demonstrating that displaying reasoning traces can indeed improve human-AI interaction. Yet, existing reasoning trace methods present unstructured text that overloads working memory \parencite{sweller1990cognitive} preventing the interactive, step-by-step engagement necessary for users to effectively question assumptions, verify logic, and provide targeted corrections. User-centric explainable AI (XAI) research demonstrates that explanations must be tailored to users' specific needs and contexts \parencite{Doshi-Velez2018}, suggesting the need for reasoning trace approaches that prioritize collaborative usability over algorithmic optimization.

This paper presents a novel approach to \textit{user-centric Chain-of-Thought reasoning} that addresses these limitations by placing interpretability and human-AI collaboration at the center of system design. Our approach generates reasoning traces as chains of self-contained, verifiable steps that are presented through an interactive user interface (UI), enabling users to effectively evaluate and correct LLM responses.
Our contributions are threefold. First, we introduce a collaborative reasoning approach that combines existing LLM reasoning methods with user-centric interpretability requirements. Through an interactive UI presenting structured reasoning traces, users can actively engage with, challenge, and correct LLM reasoning while maintaining equivalent task performance to standard CoT approaches. Evaluation on mathematical reasoning tasks demonstrates that LLMs reliably generate these user-centric reasoning traces without sacrificing solution quality.

Second, our user study reveals that structured reasoning formats significantly improve perceived usefulness and ease of use compared to standard CoT approaches, demonstrating that users feel more capable of assessing AI reasoning based on our approach to user-centric CoT. This combination of reasoning performance and user-centric interpretability represents a meaningful contribution to both the LLM reasoning and XAI literature streams.
Third, we provide a complete technical implementation demonstrating the practical feasibility of user-centric CoT reasoning. Our system encompasses prompting for generation of structured reasoning traces, robust parsing, and an interactive UI supporting cross-referencing and step-level feedback. Importantly, this design enables integrating any LLM into our approach without task-specific fine-tuning.

The remainder of this paper is structured as follows: In the next two sections, we illustrate the problem context and provide an overview of related work. We then propose our novel approach, followed by a demonstration of its applicability and an evaluation of its efficacy. Finally, we discuss implications for theory and practice, reflect on limitations, and conclude with directions for further research.

\section{Problem Context}

LLMs have achieved widespread deployment across diverse real-world domains including healthcare, finance, education, and e-commerce \parencite{raza2025industrial}. Their remarkable flexibility enables them to perform varied tasks without task-specific training through few-shot and in-context learning \parencite{brown_language_models_are_few_shot_learners}, making them valuable tools for complex cognitive tasks that require multi-step reasoning and analysis. Despite their impressive capabilities, LLMs exhibit critical limitations that constrain reliable deployment in collaborative settings. They are prone to hallucinations, producing convincing but factually incorrect or fabricated information \parencite{huang2024largelanguagemodelsselfcorrect}, and struggle with self-correction, often requiring external feedback to improve their outputs \parencite{huang2024largelanguagemodelsselfcorrect, gou2024critic}. While automated feedback tools such as compilers and interpreters have proven effective at correcting LLM outputs in code-related domains, these tools are largely absent in other domains, leaving humans responsible for verifying and correcting LLM outputs across many critical tasks.

This human oversight becomes particularly important in complex reasoning tasks, which require AI systems to draw inferences, derive conclusions, or make decisions over multiple steps. Unlike simple pattern recognition, reasoning involves manipulating knowledge representations to solve problems and justify outcomes. For instance mathematical word problems, require models to parse natural language scenarios, identify relevant operations, and execute multi-step calculations to reach solutions \parencite{cobbe2021training}. Current approaches to LLM reasoning rely on CoT prompting, where models explicitly generate intermediate reasoning steps \parencite{wei_chain_of_thought}. The resulting reasoning traces record these intermediate steps, providing insight into how an LLM reached its final solution. However, existing work optimizes reasoning approaches primarily to improve LLM's task performances rather than enhance human understanding. While initial experiments have investigated whether displaying reasoning traces can improve human-AI interaction \parencite{schelhorn2025impact}, no prior work has examined how reasoning traces should be designed specifically for effective human-AI collaboration. This shortcoming becomes critical when considering findings from user-centric XAI research, which emphasize that explanations must be tailored to users' specific needs and contexts to be effective \parencite{forster2020evaluating}. Current reasoning traces remain designed for model performance enhancement, creating a mismatch between what LLMs produce and what humans need for effective collaboration and oversight.

Cognitive Load Theory provides crucial insights for designing effective human-AI interfaces \parencite{sweller1990cognitive, xu2014nature}. The theory distinguishes between intrinsic cognitive load—the inherent complexity of the task itself—and extraneous cognitive load—unnecessary mental effort arising from poor presentation formats \parencite{paas2020cognitive}. Recent XAI research demonstrates that explanation design directly impacts users' cognitive processing, with structured presentations reducing extraneous load while maintaining the intrinsic load necessary for meaningful decision engagement \parencite{herm2023impact, weller2025improving}.
Current CoT reasoning violates these principles through continuous text presentation that forces simultaneous tracking of multiple reasoning components such as extracted facts, logical inferences, and intermediate conclusions, creating substantial extraneous load during verification—especially when errors cascade through complex tasks. This places an excessive cognitive burden on users who must mentally parse and validate interconnected reasoning steps simultaneously.

\section{Related Work}
Our user-centric CoT reasoning approach bridges LLM reasoning and XAI research, leading us to review the current state of both fields. Additionally, we review XAI evaluation methodologies to provide background for our user-centric design and evaluation approach.

\subsection{Multi-Step Reasoning in LLMs}

LLMs can solve diverse problems without task-specific training through in-context learning, where the model receives task instructions and example problem-answer pairs alongside the problem instance \parencite{brown_language_models_are_few_shot_learners}. While effective for many applications, this approach faces limitations when applied to complex reasoning tasks that require systematic problem decomposition and multiple logical steps. CoT prompting addresses these limitations by incorporating demonstration examples that model step-by-step reasoning processes \parencite{wei_chain_of_thought}. This technique improves performance across arithmetic, commonsense, and symbolic reasoning tasks by guiding the model to decompose complex problems into manageable sub-steps. CoT prompting primarily improves task performance through enhanced accuracy in final predictions, while additionally generating reasoning traces that explain the model's decision-making process \parencite{wei_chain_of_thought}. Building on the success of CoT reasoning, recent advances develop more sophisticated methodologies. These include self-consistency techniques that aggregate multiple reasoning paths \parencite{wang2023selfconsistency}, tree-of-thought approaches that explore different reasoning branches \parencite{yao2023tree}, and agent-based systems that integrate external tools and iterative refinement mechanisms \parencite{yao2023react}. These developments establish explicit reasoning trace generation as a fundamental component of modern LLM-based AI systems.

\subsection{XAI for LLMs}

XAI focuses on providing human-understandable explanations for outputs of AI systems \parencite{brasse2023explainable}. Explanations serve multiple purposes, including increasing user trust and supporting model debugging \parencite{Meske02012022}. LLMs present unique explainability challenges due to their output complexity, interactive nature, hallucination tendencies, and diverse use cases \parencite{schneider2024explainable}.
Recent XAI research for LLMs has addressed several key areas. Trust calibration and appropriate reliance studies examine how explanations, source attribution, and inconsistencies affect user reliance on LLM outputs \parencite{kim2025fostering}, while comparative research evaluates human-generated versus machine-generated explanations for understanding AI errors \parencite{pafla2024unraveling}. Hallucination detection has led to specialized interfaces that identify and highlight potentially unreliable content in LLM responses \parencite{leiser2024hill}. Research has also explored self-generated explanations, such as automated assessment of comprehension strategies \parencite{nicula2023automated}, alongside established approaches such as counterfactual explanations \parencite{cheng2024interactive}. Building on this foundation, our approach extends XAI research for LLMs by designing and evaluating explanations in the form of structured interactive reasoning traces.

\subsection{Evaluation of XAI}
Empirical findings from prior XAI studies demonstrate that explanations can fail to meet their intended outcomes. For instance, \textcite{alufaisan2021does} find that explanations had no effect on decision accuracy, while \textcite{schrills2020color} observe no effect on perceived trustworthiness. These mixed findings underscore the importance of systematic evaluation of XAI methods.
The XAI evaluation framework by \textcite{Doshi-Velez2018} provides a systematic approach to address these challenges, distinguishing three types of evaluation: functionally-grounded evaluation assesses explanations based on proxy measures without involving users, human-grounded evaluation is conducted with lay users performing simplified tasks, and application-grounded evaluation is conducted with real users in actual application settings. This multi-tiered framework is essential because each evaluation type captures different aspects of explanation quality and utility.
Explanations can only meet their objectives if they explicitly address users' specific needs for interpretability \parencite{forster2020fostering}. Researchers increasingly build on insights from the social sciences to capture universally applicable needs for interpretability \parencite{MILLER20191}. \textcite{brasse2023explainable} demonstrate how rigorous evaluation can measure how well explanations satisfy users' needs, while \textcite{forster2020fostering} show how such evaluation can guide iterative improvements to XAI methods through alternating between functionally-grounded and human-grounded evaluation. Against this background, it is essential to conduct comprehensive evaluation of XAI methods that systematically assesses explanation quality across functional, human-centered, and application-specific dimensions while explicitly accounting for users' needs and context-specific requirements.

\section{A New Approach to User-Centric Chain-of-Thought Reasoning}

We aim to design a novel approach that makes LLM reasoning traces interpretable for users \parencite{peffers2007design}. Current reasoning traces are optimized for model performance rather than user comprehension. Our design aims to improve human-AI collaboration by reducing users' extraneous cognitive load through structured presentation of reasoning components. We expand this body of knowledge by proposing three key components that constitute our approach: 1) using an answer schema with self-contained reasoning steps, 2) presenting reasoning traces in an interactive UI, and 3) generating responses in a structured output format.

\subsection{Basic Idea}

\textbf{Answer Schema with Self-Contained Reasoning Steps}. Cognitive load theory establishes that extraneous cognitive load—mental effort unrelated to the core task—impairs comprehension and decision-making \parencite{paas2020cognitive, sweller1990cognitive}. Long, unstructured reasoning traces impose such extraneous load by forcing users to mentally parse and organize information while simultaneously evaluating correctness. Our approach directly addresses this cognitive challenge by generating reasoning traces that follow an answer schema of discrete, self-contained steps. Each step includes all necessary context and information for independent verification, eliminating the need for users to maintain mental models of entire solution paths. Additionally, each step serves as a checkpoint by forming a partial solution, enabling users to preserve valid portions when correcting errors. This modular presentation allows users to focus cognitive resources on evaluating individual reasoning components rather than tracking complex dependencies across lengthy traces.

\begin{figure*}[thb]
    \centering
	\includegraphics[width=\linewidth]{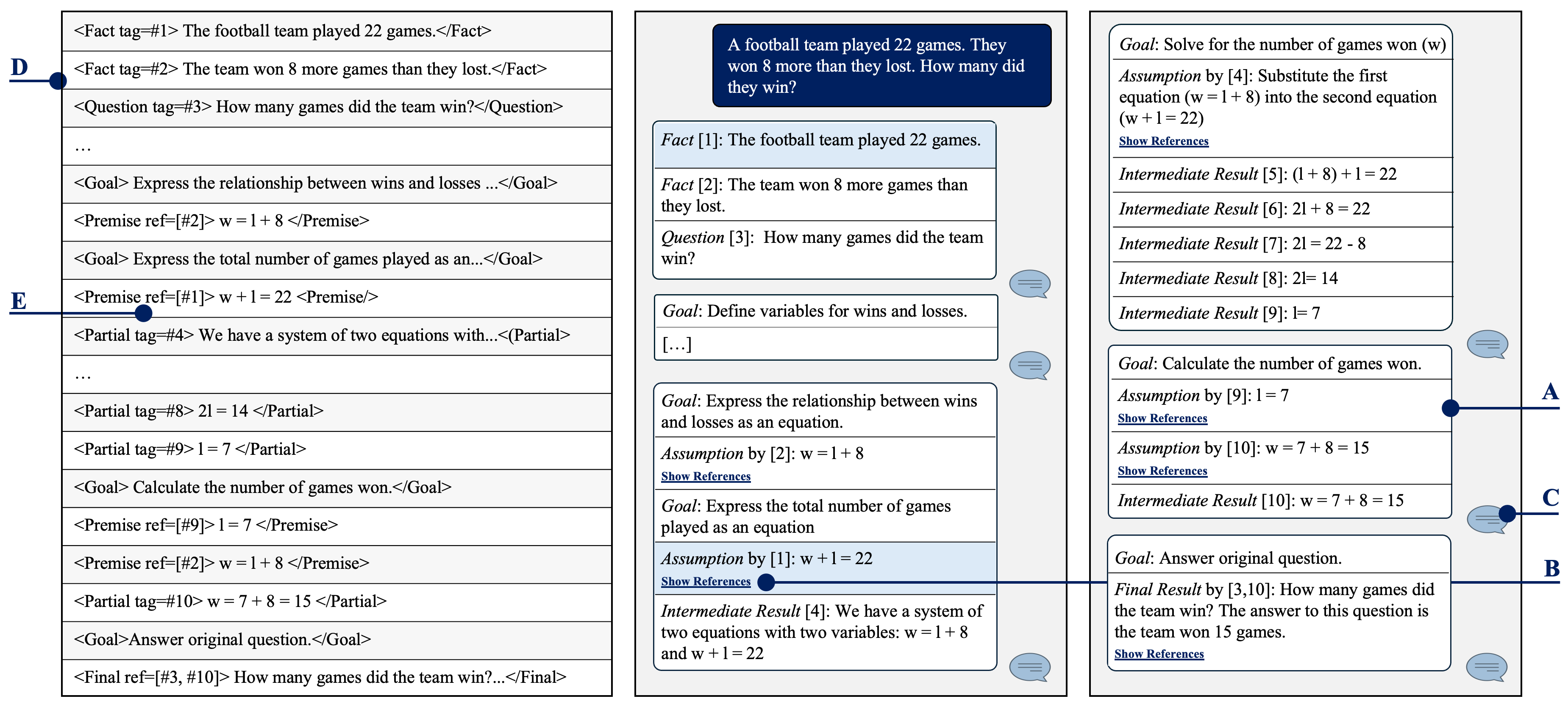}
	\caption{Example of our user-centric reasoning approach: structured, DSL-formatted reasoning trace (left) and corresponding UI implementation (middle, right).}
	\label{fig: artifact-deep-dive}       
\end{figure*}

\textbf{Interactive UI}. Research on XAI shows that effective explanations require interactive dialogue where users can probe and question AI outputs \parencite{MILLER20191}, yet current CoT approaches only allow users to act on complete reasoning traces. Our design counters this through an interactive feedback system that enables granular engagement on partial reasoning traces, decomposing the cognitive task into manageable components so users can efficiently verify individual steps rather than maintain complex mental models of entire solution paths. Reasoning steps are visually separated (Fig.\ref{fig: artifact-deep-dive}-A) and steps that depend on one another are connected via a cross-referencing mechanism. Users can click to highlight all dependencies of a segment to effectively trace information throughout the reasoning trace (Fig.\ref{fig: artifact-deep-dive}-B). On top of that users can correct mistakes in reasoning steps by providing targeted feedback on the step with the mistake (Fig.\ref{fig: artifact-deep-dive}-C). An updated answer which reuses the prior, correct steps is then created by the LLM. This transforms the interaction into a collaboration where humans work alongside AI to create a correct solution to complex reasoning problems.

\textbf{Structured LLM Output}. For our approach to achieve practical applicability, interpretability improvements cannot compromise the underlying LLM's ability to correctly answer questions. Our method achieves this balance by leveraging the LLM's existing capabilities to generate structured output. We prompt LLMs to embed XML-like tags directly within their reasoning traces, creating a unified response that contains both reasoning content and structural meta-information required for interpretability features. These semantic tags (e.g., \verb|Goal, Premise, Partial|, and \verb|Final|, cf. Fig.\ref{fig: artifact-deep-dive}-D) organize the reasoning trace while tag attributes (e.g., \verb|ref=[#2]|, cf. Fig.\ref{fig: artifact-deep-dive}-E) establish the cross-referencing relationships that power the interactive interface. 

\subsection{Technical Details}

Our approach prompts LLMs to create reasoning traces in a domain-specific language (DSL) which can be parsed into a structured format that enable interactive UI for human-AI collaboration. This section details the three stages of our technical pipeline: prompt design and LLM generation, DSL parsing and structuring, and the implementation of the interactive UI. We examine each component and data representation that facilitates the transformation from raw LLM output to collaborative reasoning experience.

\begin{figure}[thb]
    \centering
    \includegraphics[width=\linewidth]{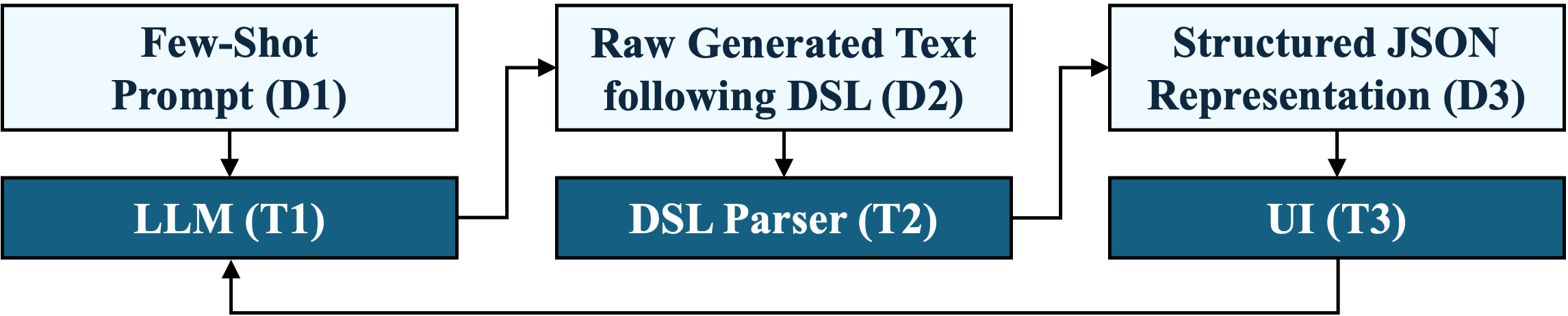}
	\caption{System architecture of the user-centric CoT approach.}
	\label{fig: instantiation}       
\end{figure}

\textbf{Prompt and Generation}. The foundation of our approach lies in guiding LLMs to generate reasoning traces that follow a specific answer schema. Within this schema, solutions are constructed step-by-step from self-contained reasoning steps, each composed of segments that serve distinct semantic purposes (cf. Fig. \ref{fig: artifact-deep-dive}-A). Reasoning steps contain segments for stating the step's goal (\verb|Goal|), the assumptions underlying the reasoning (\verb|Premise|), and intermediate results (\verb|Partial|). Beyond these core argumentative components, the schema incorporates two specialized step types: summary steps at the beginning of each trace that extract key information (\verb|Fact|) and questions (\verb|Question|) from the original problem, and final steps that present the overall goal (\verb|Goal|) alongside the definitive result (\verb|Final|). While our approach employs carefully designed few-shot prompts containing exemplar problems with complete solutions—similar to standard CoT prompting—the critical distinction lies in the structured response format (cf. Fig. \ref{fig: instantiation}-D1). Rather than generating free-form reasoning text, the LLM produces responses following a domain-specific language (DSL) that encodes both reasoning content and structural meta-information (cf. Fig. \ref{fig: instantiation}-T1). This DSL organizes each reasoning trace as a sequence of segments—the smallest meaningful units of reasoning content—using XML-like patterns that capture not only the semantic type of each segment but also carry essential meta-information for cross-referencing and navigation, as detailed in Table \ref{tab:regex-components}.

\begin{table}[ht]
\centering
\caption{Regex definition of DSL for structured reasoning step segments.}
\vspace{3pt}
\label{tab:regex-components}

\centering
\renewcommand{\arraystretch}{1}
\begin{tabularx}{\columnwidth}{l}
\toprule
\textbf{Pattern of a Reasoning Segment} \\
\texttt{\small \textbf{<}\textcolor{teal}{(?P<type>TYPE)}\textcolor{magenta}{(?P<attrs>ATTRS)}\textbf{>}} \\
\texttt{\small \textcolor{blue}{(?P<content>CONTENT)}} \\
\texttt{\small \textbf{</}\textcolor{teal}{(?P=type})\textbf{>}} \\[0.3em]
\midrule
\textbf{Named Capturing Groups} \\
\texttt{\small \textcolor{teal}{TYPE} = (Fact|Question|Goal|Premise|} \\
\texttt{\small Partial|Final)} \\
\texttt{\small \textcolor{magenta}{ATTRS} = (key=val)*} \\
\texttt{\small \textcolor{blue}{CONTENT} = .*?} \\[0.3em]
\midrule
\textbf{Example:} \\
\texttt{\small "<Premise ref=[\#2]>w=l+8</Premise>"} $\mapsto$ \\[0.2em]
\texttt{\small \textcolor{teal}{type}="Premise"} \\
\texttt{\small \textcolor{magenta}{attrs}="ref=[\#2]"} \\
\texttt{\small \textcolor{blue}{content}="w=l+8"} \\[0.3em]
\bottomrule
\end{tabularx}
\end{table}

\textbf{Parsing of LLM Response}. Following the structured prompt, the LLM generates a response in our DSL format (cf. Fig. \ref{fig: instantiation}-D2), which then undergoes systematic parsing to extract structured reasoning components (cf. Fig. \ref{fig: instantiation}-T2). Each segment follows the pattern \verb|<type attributes>content</type>| where the type indicates the segment's semantic role, attributes store meta-information such as cross-references using key-value pairs like \verb|ref=[#2,#5]|, and content contains the actual reasoning substance. The regex-based parsing approach detailed in Table \ref{tab:regex-components} can be easily implemented in most common programming languages, making our approach broadly accessible for integration into existing systems. However, since LLMs are guided by prompts rather than constrained to adhere to the DSL strictly, generated outputs may deviate from the DSL specification. To address this challenge, our parsing system incorporates error resilience mechanisms that can handle malformed inputs such as unclosed tags or missing attributes. When parsing encounters issues, graceful degradation ensures the system can display partial results or fallback to raw text presentation, maintaining usability even with imperfect LLM outputs. Finally, the successfully parsed segments are organized into coherent reasoning steps based on their semantic types and sequential relationships, while cross-reference attributes are processed to construct a navigable dependency graph that captures the logical order of the reasoning steps.

\textbf{Interactive User Interface}. The structured reasoning data (cf. Fig. \ref{fig: instantiation}-D3) serves as input for a UI (cf. Fig. \ref{fig: instantiation}-T3) that transforms static reasoning traces into an interactive collaborative user experience. The UI processes both content and meta-information including segment types, identifiers, and dependency relationships to create the user experience illustrated in Fig. \ref{fig: artifact-deep-dive}. Central to this interaction is the cross-referencing system that enables users to click on segments containing references to other segments and visualize all dependencies throughout the reasoning chain, facilitating comprehensive verification and understanding of the logical structure. Each reasoning step functions as a checkpoint that captures a partial solution state, enabling effective reuse of validated reasoning components during correction processes. When users identify errors, the correction mechanism captures their feedback and integrates it as user input within the reasoning chain. The system then generates a new LLM response using all reasoning steps up to the correction point plus the user's feedback as the regeneration prompt. This approach preserves the integrity of partial solutions by maintaining all valid reasoning steps preceding the correction as checkpoints, avoiding the computational overhead and potential error propagation associated with  the regeneration of a complete solution while enabling iterative collaborative refinement of complex reasoning traces.

\section{Demonstration and Evaluation}

As an essential component of the Design Science research methodology \parencite{peffers2007design}, we demonstrate our approach's applicability and evaluate its effectiveness in a realistic setting. Following the framework for XAI evaluation proposed by \textcite{Doshi-Velez2018}, we conduct a series of evaluations. First, we conduct a functionally-grounded evaluation to assess whether our user-centric CoT approach compromises LLM task performance (section \ref{subsection:functional-eval}). Second, we employ human-grounded evaluation through a user study examining both subjective constructs and behavioral measures to determine whether user-centric CoT reasoning enhances human-AI collaboration (section \ref{subsection:human-grounded-eval}).

\subsection{Demonstration}

\textbf{Setting}. Rigorous demonstration and evaluation require experimental settings where users' tasks and IT systems closely resemble real-world conditions \parencite{peffers2007design}. We selected mathematical word problems as our demonstration domain, specifically having users collaborate with an LLM-based chat system to solve problems from the GSM8K dataset \parencite{cobbe2021training}. To ensure data integrity and avoid potential training data contamination, we exclusively used the test split containing 1,319 previously unseen questions. Mathematical word problems serve as prototypical reasoning tasks, making them highly relevant for evaluating human-AI collaboration across diverse analytical domains.

\textbf{Instantiation}. We instantiated our approach in a web application with three components (cf. Fig. \ref{fig: instantiation}). The generation system (T1) employs an LLM with a one-shot prompt (D1) containing a GSM8K example formatted in our XML-like DSL to generate solutions with tagged, cross-referenced reasoning segments. We evaluate this generation approach across three open-weights models of different sizes: Llama3.2-3B (small), Gemma2-27B (medium), and Llama3.3-70B (large). The regex-based parser (T2) processes the DSL-formatted response (D2) into structured JSON (D3), extracting tag names, attributes, and content while maintaining cross-reference relationships and providing error recovery for malformed structures. The React-based web UI (T3) renders reasoning traces as separated atomic elements with clickable cross-references and step-level feedback mechanisms. User feedback triggers targeted repair by passing corrections and prior steps back to the LLM for selective regeneration of subsequent steps while preserving valid preceding reasoning links.

\subsection{Functional Evaluation}
\label{subsection:functional-eval}
We conducted a functional evaluation following XAI framework put forward by \textcite{Doshi-Velez2018} to verify that generating user-centric reasoning traces preserves model performance on the underlying task.
We compared our user-centric CoT approach against standard CoT reasoning on the GSM8K test set across three model sizes (small, medium, and large), measuring both accuracy and response length (cf. Table \ref{tab:model_comparison}). The evaluation reveals that medium and large models maintain equivalent high-level performance when generating user-centric reasoning traces compared to standard CoT approaches, demonstrating that our structured format does not degrade reasoning capabilities. Notably, our approach enhanced performance for the small LLMs which exhibited poor baseline performance.
Our user-centric CoT approach consistently produced responses approximately three times longer than standard CoT across all model sizes. Even accounting for structural markup (XML-like tags, cross-references, meta-information), the underlying reasoning remains more verbose, as self-contained steps necessarily include context that standard CoT leaves implicit.

\begin{table}[htb]
\caption{Accuracy (Acc.) and mean output length (Len., in tokens) for standard and user-centric CoT prompting on GSM8K.}
\vspace{3pt}

\centering
\setlength{\tabcolsep}{4pt}
\renewcommand{\arraystretch}{1.1}
\begin{tabularx}{\columnwidth}{lcc|cc}
\toprule
\textbf{Model} & \multicolumn{2}{c|}{\textbf{User-Centric CoT}} & \multicolumn{2}{c}{\textbf{Standard CoT}} \\
 & \textbf{Acc.} & \textbf{Len.} & \textbf{Acc.} & \textbf{Len.} \\
\midrule
Llama3.3-70b & 0.959 & 1088.5 & 0.953 & 345.1 \\
Gemma3-27b   & 0.943 & 1114.6 & 0.941 & 301.2 \\
Llama3.2-3b  & 0.487 & 1172.0  & 0.367 & 338.1 \\
\bottomrule
\end{tabularx}
\label{tab:model_comparison}
\end{table}

\subsection{Human-Grounded Evaluation}
\label{subsection:human-grounded-eval}

\begin{figure*}[thb]
    \centering
	\includegraphics[width=\linewidth]{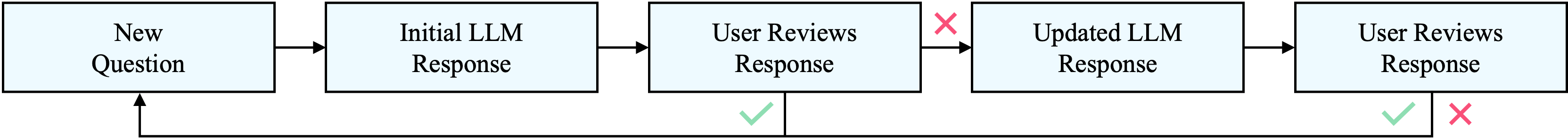}
	\caption{Procedure of the main task in the user study.}
	\label{fig: study-procedure}       
\end{figure*}

We evaluated the effectiveness of our proposed approach in helping users verify and correct complex LLM outputs in an online experiment \parencite{fink2022and}. This corresponds to human-grounded evaluation in the XAI evaluation framework put forward by \textcite{Doshi-Velez2018}. First and foremost, we aimed to evaluate the effectiveness of our approach compared to a competing approach \parencite{peffers2007design}, i.e., LLM responses based on standard CoT prompting \parencite{wei_chain_of_thought}. In the following, we describe the study design, study procedure, measurement and
analysis, participants, and results.

\textbf{Evaluation Objective}. Within this collaborative problem-solving context, the primary objective of XAI is to enhance users' ability to evaluate and effectively partner with the AI system \parencite{Meske02012022}. The user-centric reasoning traces in our approach aim to not merely justify the AI's conclusions, but to provide meaningful support that enables users to critically assess the system's reasoning process and leverage it as a reliable collaborative partner in solving complex analytical tasks.

\textbf{Experimental Design and Procedure}. To evaluate whether user-centric reasoning traces effectively enhance interpretability and human-AI collaboration, we designed a between-subjects online experiment. The treatment group received answers generated using our user-centric CoT approach, while the control group received answers generated using standard CoT reasoning. Participants served as AI collaborators, working with a chat-based AI system to solve math word problems by verifying the AI's responses and providing corrections when necessary.\footnote{Screenshots from the online experiment, LLM responses, and examples of corrections submitted by participants are available at \url{https://doi.org/10.17605/OSF.IO/EFAZ7}.} The experiment was conducted with Llama3.3-70B as the underlying LLM.

Our study comprised three parts: an introduction, the main task, and a post-survey. Participants engaged in a structured evaluation consisting of an introductory session followed by four main task rounds and a concluding survey measuring subjective perceptions.
In the introduction, participants received information about their task—checking if an AI-generated solution to a mathematical word problem is correct and providing textual feedback to fix possible mistakes. After completing an example task to familiarize themselves with the system, participants proceeded through four rounds of human-AI collaborative problem-solving using GSM8K mathematical word problems \parencite{cobbe2021training}. To incentivize participants to pay close attention and provide genuine feedback, we informed them about attention checks throughout the study and offered bonus compensation for the top 25\% of participants based on performance in the main task.

Each round followed a standardized protocol (cf. Fig. \ref{fig: study-procedure}): (1) the system presented a GSM8K question with an LLM-generated solution, (2) the participant evaluated the AI's response for correctness, and (3) the system proceeded based on the participant's assessment. For responses deemed correct, the session advanced to the next question. For responses identified as incorrect, participants provided targeted feedback at the specific reasoning step where they detected the error, triggering regeneration of subsequent steps, after which participants re-evaluated the updated response before proceeding.
The task set was deliberately balanced, with each participant encountering two problems that Llama3.3-70B had solved correctly and two it had solved incorrectly, presented in randomized order. Participants in the treatment group received solutions generated using our user-centric approach, whereas control group participants received solutions generated using the standard CoT prompt from \cite{wei_chain_of_thought}. Following completion of all four rounds, participants answered a post-task survey including subjective perception measures and two attention checks.

\textbf{Measurement}. To evaluate whether our approach improves human-AI interaction when users check and correct complex LLM responses, we measured both user perceptions and task performance.
For user perceptions, we employed established constructs from  XAI and IS literature: \textit{trust} \parencite{Meske02012022}, \textit{usefulness}, and \textit{ease of use} \parencite{davis1989perceived}, all measured on 7-point Likert scales (1 = strongly disagree, 7 = strongly agree).
For task performance, we assessed two behavioral outcomes: \textit{judgment accuracy}—the proportion of initial LLM answers correctly classified as right or wrong—and \textit{correction likelihood}—the share of incorrect initial answers where user feedback led to a correct updated response.

\textbf{Participants}. We recruited 122 participants from Prolific, requiring English fluency and a minimum high school education. We excluded 25 participants who failed attention checks and 36 who copied text from math questions or LLM answers at least twice (participants received warnings after the first detected copy to prevent external tool usage). The final sample comprised 80 participants: 37 in the treatment group and 43 in the control group. Participants had a median age of 32 years; 45 identified as male, 34 as female, and one did not specify gender.

\begin{figure}[thb]
    \centering
	\includegraphics[width=\linewidth]{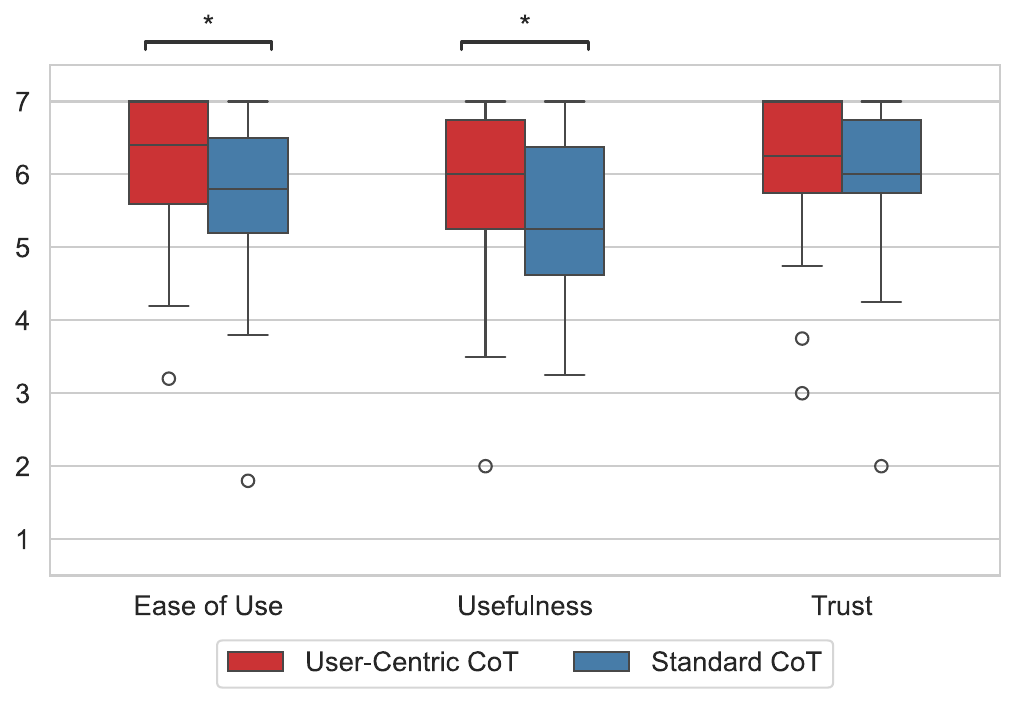}
	\caption{Comparison of Treatment (User-Centric CoT) and Control (Standard CoT) regarding subjective constructs.}
	\label{fig: construct-boxplot}       
\end{figure}

\textbf{Analysis and Results}. We calculated construct scores for each participant by averaging the respective item scores.
Internal consistency reliability was established for all constructs, with Cronbach’s alpha values of $0.95$ for \textit{ease of use}, $0.95$ for \textit{usefulness} and $0.88$ for \textit{trust}, all exceeding the recommended threshold of $0.7$ \parencite{gefen2000structural} without having to remove any item. To compare the treatment and control group regarding each subjective construct, we first tested the construct score for normality. As the construct scores were not normally distributed (Shapiro-Wilk test, $p < 0.001$ for all construct scores). we conducted one-sided Mann-Whitney U tests for the hypothesis that construct scores in the treatment group are higher than in the control group. Participants in the treatment group provided significantly higher values for \textit{ease-of-use} ($U=1004.0$, $p<0.05$) and \textit{usefulness} ($U=970.5$, $p<0.05$). Differences for \textit{trust} were not significant. Fig. \ref{fig: construct-boxplot} visualizes the distribution of construct scores. Our analysis for the behavioral measures \textit{judgment accuracy} and \textit{correction likelihood} was carried out analogously. Both measures were not normally distributed (Shapiro-Wilk test, $p<0.001$ for both measures) and two-sided Mann-Whitney $U$ tests did not show significant differences between the groups for both measures. Table \ref{tab:compact-descriptives} provides an overview of the empirical results.

\begin{table}[ht]
\centering
\setlength{\tabcolsep}{3pt}
\renewcommand{\arraystretch}{1.1}
\caption{Descriptive statistics by group. $N$ is smaller for the Correction Likelihood because some participants never correctly recognized a mistake in the initial answer.}
\vspace{3pt}
\begin{tabularx}{\columnwidth}{llr|lr}
\toprule
\textbf{Measure} & \multicolumn{2}{c|}{\textbf{User-Centric CoT}} & \multicolumn{2}{c}{\textbf{Standard CoT}} \\
& \textbf{Mean$\pm$SE} & \textbf{$N$} & \textbf{Mean$\pm$SE} & \textbf{$N$} \\
\midrule
\multicolumn{5}{l}{\textit{Subjective Constructs}} \\
Ease of Use   & $6.09 \pm 0.17$ & 37 & $5.69 \pm 0.16$ & 43 \\
Usefulness    & $5.83 \pm 0.19$ & 37 & $5.43 \pm 0.18$ & 43 \\
Trust         & $6.13 \pm 0.16$ & 37 & $6.03 \pm 0.14$ & 43 \\
\addlinespace
\multicolumn{5}{l}{\textit{Behavioral Measures}} \\
Judgment Acc.     & $0.62 \pm 0.04$ & 37 & $0.68 \pm 0.04$ & 43 \\
Correction Lik.    & $0.36 \pm 0.13$ & 13 & $0.60 \pm 0.08$ & 24 \\
\bottomrule
\end{tabularx}
\label{tab:compact-descriptives}
\end{table}

\section{Implications, Limitations and Future Research}

Our findings demonstrate that user-centric CoT reasoning significantly improves perceived usefulness and ease of use compared to standard CoT. While we observed no significant differences in perceived trust and the behavioral measures judgment accuracy, correction likelihood, the empirical results suggest that structured explanations help users feel more capable of assessing AI reasoning.

While our study provides valuable insights, limitations should be noted. Our user-centric approach improved perceived usefulness and ease of use but not the behavioral measures judgment accuracy and correction likelihood, possibly due to considerably longer explanations \parencite{lombrozo2007simplicity} that may offset cognitive benefits—a deliberate design trade-off where we prioritized self-contained, verifiable reasoning steps over brevity (cf. Table 2). The absence of trust differences aligns with \textcite{jacovi2021trust} findings that trust formation requires domain expertise and genuine vulnerability—conditions difficult to establish in an online experiment with low-stakes mathematical problems.

Our demonstration and evaluation focused exclusively on mathematical word problems, representing only one specific reasoning domain. Future research should extend this approach to other reasoning tasks such as logical inference, scientific problem solving, and tool-use scenarios where AI systems must coordinate multiple capabilities. Each domain may require domain-specific semantic tags and interaction patterns, providing opportunities to test the generalizability of our DSL-based approach.
 
The empirical findings should be interpreted considering our sample of $N=80$ online participants, who may differ from professional users in domain expertise and engagement. Additionally, unmeasured individual differences in cognitive style, domain expertise, and AI literacy may have influenced our results. Future research should involve larger-scale studies with domain experts evaluating reasoning traces on real tasks, systematically measuring and controlling for relevant user characteristics. Future work should also directly measure cognitive load to empirically validate whether our approach reduces users' mental effort.

\printbibliography

\end{document}